\documentclass{elsart}
\usepackage{epsfig}
\usepackage{amssymb,latexsym,amsfonts,amsmath}
\begin{document}
\begin{frontmatter}
\title{Effects of epidemic threshold definition on disease spread statistics}
\author[unmp]{C. Lagorio},
\author[unmp]{M. V. Migueles},
\author[unmp,cps]{L. A. Braunstein},
\author[oxford]{E. L\'opez},
\author[unmp]{P. A. Macri}.
\address[unmp]{Instituto de Investigaciones F\'isicas de Mar del Plata (IFIMAR)-Departamento 
de F\'isica, Facultad de Ciencias Exactas y Naturales, Universidad Nacional de Mar del 
Plata-CONICET, Funes 3350, (7600) Mar del Plata, Argentina.}
\address[cps]{Center for Polymer Studies, Boston University, Boston, Massachusetts 02215, USA}
\address[oxford]{CABDyN Research Cluster, Physics Department, and Sa\"{\i}d
Business School, University of Oxford, Park End Street Oxford, OX1 1HP, United Kingdom}

\begin{abstract}
We study the statistical properties of the SIR epidemics in 
heterogeneous networks, when an epidemic is defined as only those SIR propagations
that reach or exceed a minimum size $s_c$.
Using percolation theory to calculate the average fractional size 
$\langle M_{\rm SIR}\rangle$ of an epidemic, we find that the strength of the 
spanning link percolation cluster $P_{\infty}$ is an upper bound to 
$\langle M_{\rm SIR}\rangle$. For small values of $s_c$, $P_{\infty}$ is no longer 
a good approximation, and the average fractional size has to be computed directly. 
The value of $s_c$ for which $P_{\infty}$ is a good approximation is found
to depend on the transmissibility $T$ of the SIR.
We also study $Q$, the probability that an SIR propagation reaches
the epidemic mass $s_c$, and find that it is well characterized by percolation
theory.
We apply our results to real networks (DIMES and Tracerouter) to measure the
consequences of the choice $s_c$ on predictions of average outcome sizes of computer
failure epidemics.
\end{abstract}

\end{frontmatter}

The study of disease spread has seen renewed interest
recently~\cite{Pastor-Satorras,Newman,Grassberger} due the
emergence of new infectious lethal diseases such as AIDS and SARS~\cite{Anderson,Colizza}.
New tools, ranging from powerful computer models~\cite{Eubank} to
new conceptual developments~\cite{Pastor-Satorras,Sander,Eben,Cohen,Lopez,Miller},
have emerged in hopes of understanding and addressing the problem effectively.

Among the new tools that have become available to tackle infectious disease
propagation, complex network theory~\cite{Barabasi-Albert,Albert} has seen considerable
interest~\cite{Colizza,Newman}, as a way to address the shortcomings
of more classic approaches~\cite{Anderson} where all individuals in the
population of interest are assumed to have an equal probability to infect all other
individuals (random-mixing).
In contrast to the random-mixing approach, complex networks (heterogenous mixing) assume that
each individual (represented by a node) has a defined set of contacts
(represented by links) to other specific individuals (called neighbors),
and infections can be propagated only through these contacts.
This new technical framework has produced novel insights that are
expected to help considerably in the fight against infectious
diseases~\cite{Cohen,Colizza}.

The use of complex network theory requires a few pieces of information
in order to be correctly applied. First, it is important to understand the kind of disease
being considered, as this will dictate the specifics of the network model that
needs to be used. For example, the flu virus usually spreads among people
that come in contact even briefly, leading to networks with fat-tailed distributions
of connections with large average degree~\cite{Eubank}. On the other
hand, sexually transmitted diseases are better described by more sparse, and fairly
heterogeneous contact networks~\cite{Anderson}. Thus, these two examples
easily illustrate one of the complications of the problem: the structure of the network
to be used. Other aspects involve the life cycle of the pathogen, seasonality, etc.
Additionally, social and practical aspects involving public health policy and strategic
planning play important roles in the problem.

Regarding the issue of network structure, a few models have been proposed
as useful substrates for disease propagation. Among these, truncated scale-free
network structures~\cite{Newman} have received considerable interest~\cite{Eben,Miller}.
In these networks, each node has a probability $P(k)$ to have $k$ links (degree $k$)
connecting to it, with $P(k)$ being characterized by the form
\begin{equation}
P(k)= \left[
k^{-\lambda}\exp(-k/\kappa)\right]/\left[\mbox{Li}_{\lambda}(e^{-1/\kappa})\right],
\label{PdK}
\end{equation}
with $k \ge k_{min}$, where $k_{min}$ is the lower degree that a node can have
and $\kappa$ is an arbitrary degree cutoff reflecting the properties of the
substrate network for the disease~\cite{fn-polylogarithm}. The reason for including the exponential cutoff is two-fold: first many real-world graphs appear to show this cutoff; second it makes the distribution normalizable for all $\lambda$, and not just $\lambda \geq 2$ ~\cite{Newman_RG}.

Another important issue of propagation relates to the
type of disease being considered and its dynamics.
In this sense, a general model for a number of diseases (including the
ones mentioned at the beginning) is the SIR model, which
separates the population into three groups: susceptible, infected and
recovered (or removed), approximating well the characteristics
of many microparasitic diseases~\cite{Anderson}. The solution
to the SIR model corresponds to the determination of the number of
susceptible, infected, and recovered individuals at a given time.
Public health officials are particularly interested in the final outcome
of the disease propagation, measured through the number of individuals
$S_{\rm SIR}$, out of a population of $N$, that became infected at any time.
Another useful way to express the solution of the model is through the average
fraction of infected individuals $\langle M_{\rm SIR} \rangle = \langle S_{\rm SIR}/N \rangle$, where $\langle \rangle$ denotes averages over realizations.

A number of details related to SIR determine the methods that correctly
yield $S_{\rm SIR}$~\cite{Eben,Miller}. One common formulation of SIR assumes that
on each time step, an infected node has a probability $\beta$ to infect
any of its susceptible neighbors, and once infected the node recovers
in exactly $t_R$ time steps. This yields an overall probability $T$, called the
transmissibility, to use any given network link of a node that becomes infected.
For this case, when the networks have very simple structure~\cite{fn-uncorrelated-networks},
$\langle M_{\rm SIR} \rangle$ can be determined using a mapping to the link percolation
model~\cite{Grassberger,Newman} of statistical physics~\cite{Stauffer} (see below). If the SIR
propagation details change, modified forms of percolation may be used~\cite{Eben,Miller}.

From the standpoint of public health policy and strategic planning,
an important technical point is how to ``define'' what is considered to be
an epidemic, because such definition determines the level of
reaction that health organizations (e.g., World Health Organization) will apply
in dealing with a particular infectious disease event.
In real-world disease spread situations, as pointed out in several
references~\cite{Newman,Eben,Miller}, epidemiologist
are obliged to define a minimum number of people infected, or threshold $s_c$
to distinguish between a so called outbreak (a small number of individuals where no large
intervention is called for), and an epidemic (a significant number of individuals in
the population requiring large scale intervention).
In Refs.~\cite{Newman,Eben,Miller}, for instance, $s_c$ has been used, but its impact on
average predictions of SIR has not been systematically addressed, even though it
is representative of the sensitivity, or urgency, that epidemiologist assign
to the disease in question.

In this paper we address the importance of $s_c$ for SIR in complex networks.
Using link percolation, we first concentrate on calculating the average fraction
size $\langle M_{\rm SIR}(T,s_c)\rangle$ over SIR model realizations for which
$S_{\rm SIR}\geq s_c$. This quantity is important in the public health community to
determine the average expectation value for the epidemic size that can arise 
given the particular pathogen and society affected, 
{\it and the epidemic threshold $s_c$ chosen}.
To calculate SIR through link percolation, we find that a reweighting procedure is necessary,
that has been previously ignored. Once this reweighting is done,
$\langle M_{\rm SIR}(T,s_c)\rangle$ for large $s_c$ approaches $P_{\infty}(T)$,
corresponding to the average fractional size of the largest percolation cluster
at $T$, but for $s_c$ smaller than a value that depends on the topology of the network, we find that $\langle M_{\rm SIR}(T,s_c)\rangle < P_{\infty}(T)$, for $T < 1$, 
indicating that the percolation result for $P_\infty$ is an upper bound. Since the choice of
$s_c$ determines what is defined to be an epidemic, we also determine $Q \equiv Q(T,s_c)$,
the probability that an SIR realization reaches $S_{\rm SIR}\geq s_c$.
Extending our results to situations such as computer networks,
where one should be able to declare an epidemic even if few computers are infected due
to the ``similarity'' of the world population of computers (i.e. sharing the same operating
system), and thus have large susceptibility, we find that similar results apply.

The rest of the article is structured as follows. Section \ref{models} introduces details
of the network model and where it applies, the link percolation method used to solve the SIR model, and the details of the reweighting procedure necessary to obtain correct
averages. Sections \ref{results} and \ref{real-nets} introduce and explain the results of
the application of the model to disease propagation events in simulated networks and
real-world examples (computer networks). Finally, Sec.~\ref{conclusions} summaries
the results of the paper and presents our conclusions.

\section{Models and algorithm}
\label{models}

To construct networks of size $N$ we use the Molloy-Reed algorithm~\cite{algorithm},
and apply it to the degree distribution given by Eq.~(\ref{PdK}). 
Simulations for this type of network have been
performed before in Refs.~\cite{Newman} and \cite{Eben} for $N=10^4$ and $10^5$, $\lambda=2$,
$k_{min}=1$, $\kappa = 5,10,20$ and $s_c=100$ and 200~\cite{Newman_Private}. 
We perform our simulations for many values of $\kappa$ but we present our results only for $\kappa=10$.
Our main results also hold for other degree distributions. Due to
the fact that the lower degree is $k_{min}=1$ ~\cite{COHENchapter} and $\kappa$ is small, the network
is very fragmented and the size of the initial biggest connected cluster (GC), labeled
here as $N_{GC}$, is typically $60\%$ of the network (for $\kappa=10$). 
For all our simulations we work only on the GC of the original network
because we are only concerned with the disease spread on connected communities. Isolated
clusters cannot propagate a disease.

To simulate SIR, we chose one node at random on the GC of the substrate network,
and infect it. Per time step, this infected node has a probability $\beta$
to infect its first neighbors. Once a neighbor has been infected, it can infect one of
its own susceptible neighbors, but it cannot be infected again nor infect another 
already infected or recovered node. All infected nodes recover after 
$t_R$ time steps of becoming infected ~\cite{fn-time}. The transmissibility $T$
is the overall probability that a node infects one of its susceptible neighbors within
the time frame $t=1$ to $t_R$,
given by $\sum_{t=1}^{t_R}\beta(1-\beta)^{t-1}=1-(1-\beta)^{t_R}$.
For every realization of SIR, the total number of nodes that become infected after
the infectious transmission has ended is given by $S_{\rm SIR}$.
The values of $S_{\rm SIR}$ satisfy a distribution $\Phi(S_{\rm SIR})$.

As mentioned in the introduction, another way to calculate $S_{\rm SIR}$ is
through the use of link percolation. This is a process in which an initial network is
modified by removing a fraction $1-T$ of its links (we use $T$ as the probability for a link
to be present because of the mapping between link percolation and our SIR model).
The effect of the removal is to generate a multitude of clusters, each being a group
of nodes that can be reached from each
other by following a sequence of edges connected to those nodes. Link percolation
has a threshold value $T=T_c$ (the percolation threshold),
characterized by the fact that, for $T<T_c$, the
size of the largest cluster typically scales as $\log N$, and for $T>T_c$, a large
cluster emerges with a size that scales linearly with $N$, alongside a number of
small clusters. Thus, a so-called percolation transition occurs at $T=T_c$ that
takes the network from disconnected to connected. In general
terms, a similar situation occurs in SIR, where a high likelihood of
transmission of the disease (large $T$) between neighbors typically leads to a large epidemic,
but if this likelihood is low (small $T$), only small localized outbreaks appear (a detailed
description of the relation is developed below).

To perform link percolation, we begin in the GC of the substrate network, and
randomly eliminate links with probability $1-T$. Each realization of this process
yields multiple connected clusters of various sizes. Realizations are then repeated
multiple times, and a distribution of cluster sizes $\phi(S_{p})$ emerges.
For the quantity $P_{\infty}(T)$, we average over the largest cluster size produced
in each realization.

The relation between SIR and link percolation can be concretely explained
in the following way: each SIR realization begins with a randomly chosen node
of the GC, and the infection propagates to a set of nodes $S_{\rm SIR}$ that can all be
traced back to the original infection. The links used in this SIR realization, on average,
where used with probability $T$ and not used with probability $1-T$. To draw the
correct connection to link percolation, we first must realize that in a given realization
of percolation, only one of the many connected clusters can be chosen to
represent the infection of SIR. By analogy with the classic Leath algorithm~\cite{Leath}
of cluster creation in percolation,
we can conclude that the clusters are randomly picked, with probability
proportional to their size $S_{p}$. Thus, one expects that the average size of SIR
realizations is equivalent to a weighted average of percolation realizations, where
the weight is given by $S_{p}$.

With the previous arguments in mind, and given the dependence of the
problem on both $T$ and $s_c$, we compute $\langle M_{\rm SIR}(T,s_c)\rangle$ 
through~\cite{fn-allclusters}
\begin{equation}
\langle M_{\rm SIR}(T,s_c)\rangle =\sum_{S_{\rm SIR}\geq s_c} \frac{S_{\rm SIR}}{N_{GC}} \; \Phi(S_{\rm SIR}).
\label{MSIR}
\end{equation}
In order to compare this to link percolation, we perform a weighted average
to obtain $\langle M_p(T,s_c)\rangle$, given by
\begin{equation}
\langle M_p(T,s_c)\rangle =\frac{\sum_{S_{p}\geq s_c} (S_{p}^2 /N_{GC})  
\; \phi(S_{p}) \;} {\sum_{S_{p}\geq s_c}S_{p} \; \phi(S_{p})}.
\label{Mp}
\end{equation}
We expect that both averages converge to the same value when enough
realizations are performed. Additionally, as $s_c$ is increased, we expect 
$\langle M_p(T,s_c \gg ~ 1) \rangle \rightarrow P_{\infty}(T)$ for $T>T_c$, because a progressively
smaller number of small clusters enters into the averaging, and only the largest
clusters are used. This creates an interesting scenario, in which $P_{\infty}(T)$
is a good approximation of the epidemic size only in the limit of a large
threshold $s_c \ge S_p^{\times}$ (a function of $T$ only, defined below), but for smaller 
$s_c$, which is important in more aggressive diseases, only 
$\langle M_p(T,s_c)\rangle$ is the correct average.

\section{Results on the relative average size of the disease}
\label{results}
\subsection{Mapping between the average fraction size using SIR simulations and
the average fraction size of all percolation cluster}

\begin{figure}[h,b]
\begin{center}
\includegraphics[width=12cm,height=10cm,angle=0]{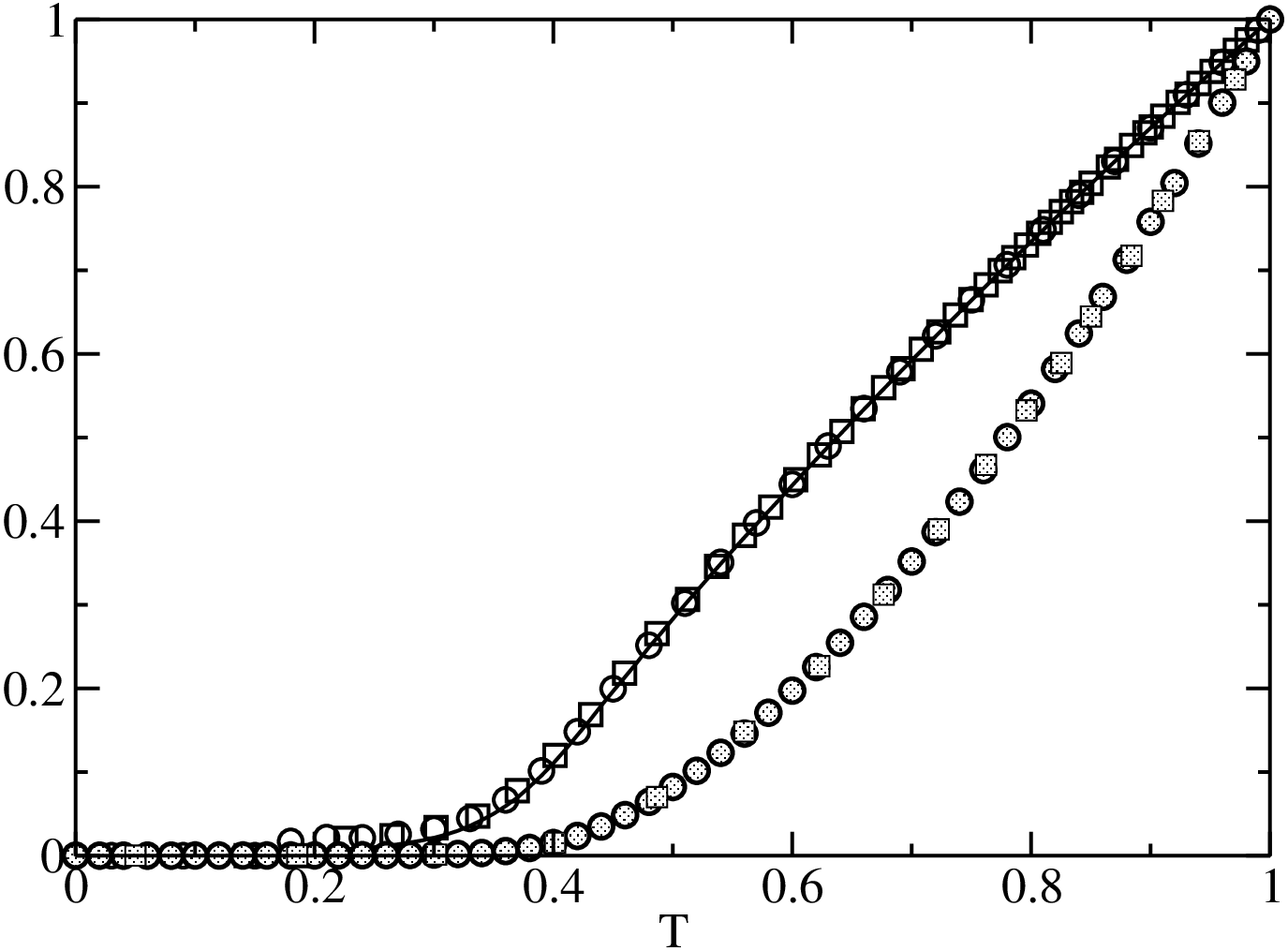}
\caption{Comparation between $\langle M_{\rm SIR}(T,s_c)\rangle$ $(\Box)$,
$\langle M_p(T,s_c)\rangle$ ($\bigcirc$), and $P_\infty(T)$ of link percolation (full line).
Empty symbols correspond to $s_c = 100$, and dotted symbols to $s_c = 1$. For
the transmissibility in the SIR problem, we used $\beta=0.05$ and a set of
values of the recovery $t_R$ to cover a wide range of $T$. All the
simulations were performed on the GC of networks with $\lambda = 2$ , $\kappa = 10$,
$k_{min} = 1$, and averaged over $10^4$ realizations. \label{fig.1}}
\end{center}
\end{figure}

As a first step, we test that indeed $\langle M_{\rm SIR}(T,s_c)\rangle$ and
$\langle M_p (T,s_c)\rangle$ are equal. In Fig.~\ref{fig.1} we plot 
$\langle M_{\rm SIR}(T,s_c) \rangle $ and
$\langle M_{p}(T,s_c) \rangle$ to check their agreement.
The two curves overlap indicating that
the mapping between the two quantities is correct. In the reminder (unless explicitly
stated), we perform our simulations using link percolation as opposed to SIR.

The mapping between the steady state of SIR and link percolation is computationally
very convenient for several reasons. First, performing simulations of SIR models is
computationally more costly than link percolation. This is due to the fact that for
SIR, only a single propagation occurs per realization, as opposed to multiple clusters
that appear for link percolation. Additionally, SIR propagation has to be performed
in a dynamic fashion, which makes it necessary to test over time a given propagation
condition, something that does not occur for link percolation, accelerating further
the simulations. Finally, this mapping is convenient because it gives another
conceptual framework in which to understand the relation between
these two problems of disease propagation and percolation models.

A final feature of Fig.~\ref{fig.1} is the plot of $P_{\infty}(T)$. This
curve displays good agreement with $\langle M_{\rm SIR}(T,s_c)\rangle$ for
the larger $s_c$. We discuss this issue further in the next subsection.

\subsection{Effects of $s_c$ on the average size of epidemics}

\begin{figure}[h,t]
\begin{center}
\includegraphics[width=9cm,height=7cm,angle=0]{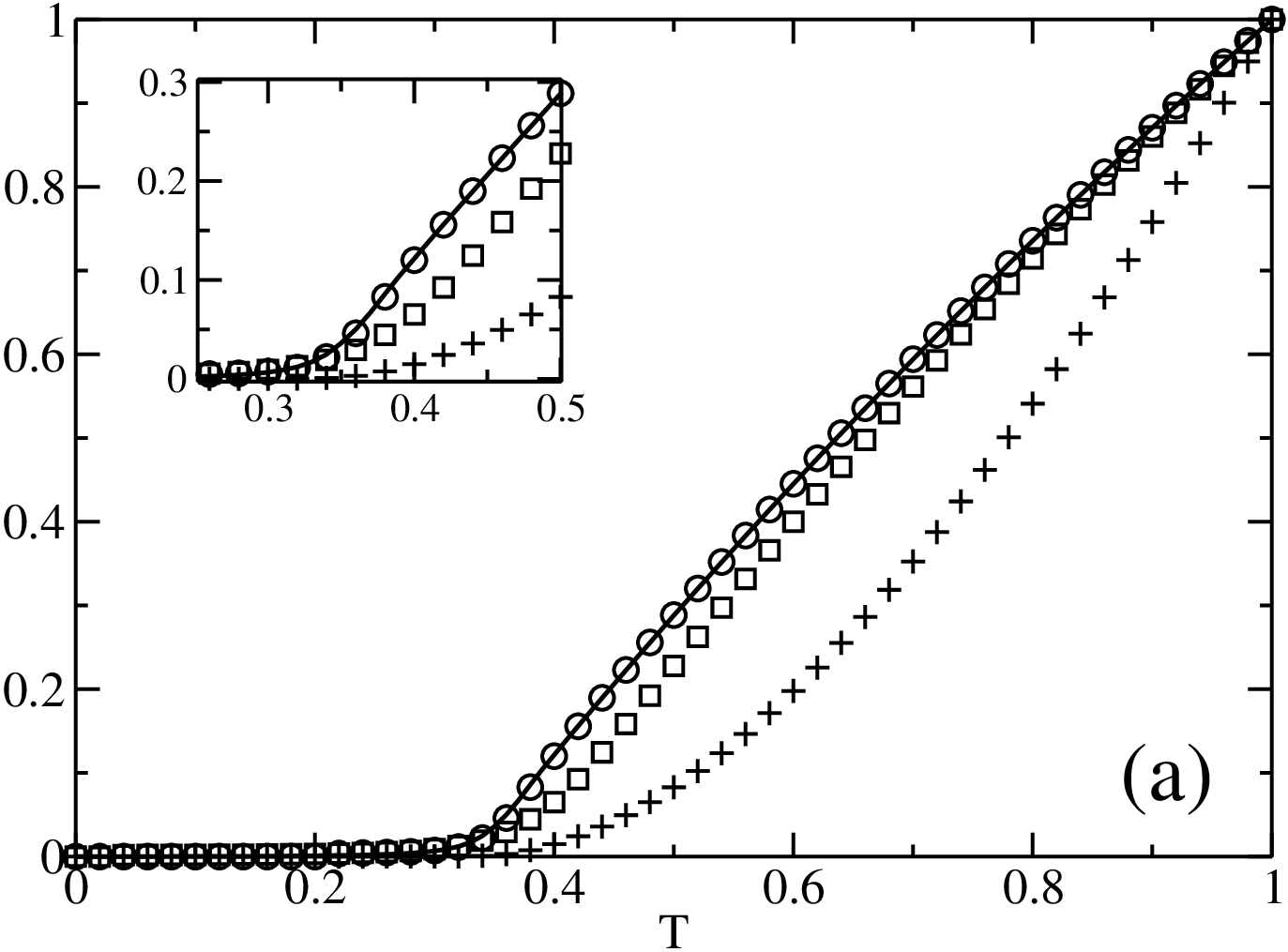}\\
\vspace{2cm}
\includegraphics[width=9cm,height=7cm,angle=0]{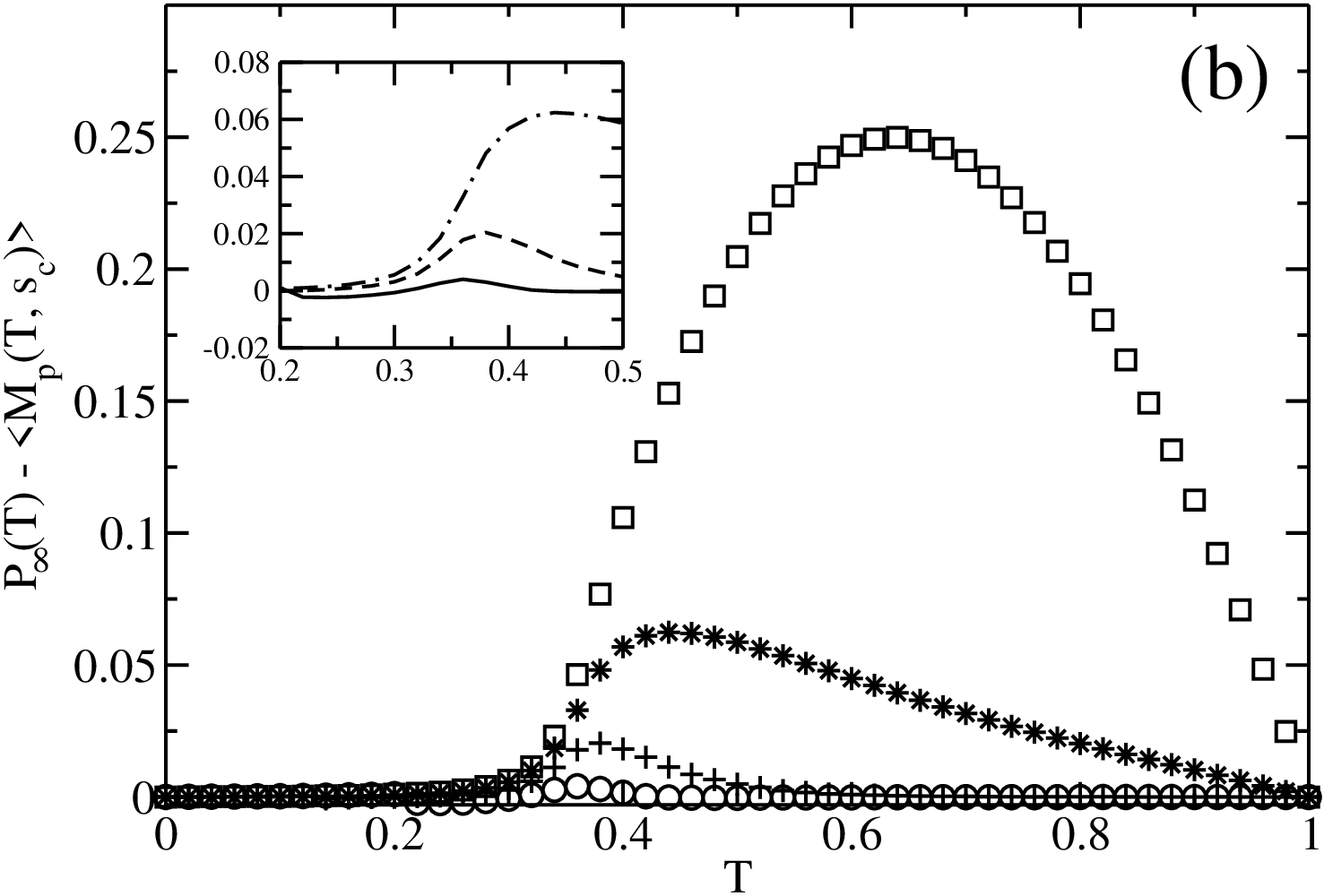}
\caption{\textbf{a)} Plot of $\langle M_p(T,s_c)\rangle$ as a function of $T$, 
for $s_c=200$ ($\bigcirc$), $s_c=10$($\Box$) and $s_c=1$ ($+$). The full line represents
$P_\infty(T)$. The inset shows the
details of the main plot close to $T_c \approx 0.32$, i.e, for $T$ near
the percolation threshold. We can observe that the departure
between $P_\infty(T)$ and $\langle M_{\rm SIR} (T,s_c) \rangle$ is not negligible.
\textbf{b)}$ P_{\infty} - \langle M_p(T,s_c)\rangle $ as a function of $T$, for $s_c=1$ 
($\Box$), $s_c=10$($\ast$), $s_c=50$ ($+$) and $s_c=200(\bigcirc)$. In the inset we plot 
the details of the main plot around $T_c$ for $s_c=10$ (dot dashed line), $s_c=50$ 
(dashed line) and $s_c=200$ (full line). We observe that $P_\infty(T)$ is an upper bound 
for $\langle M_p(T,s_c)\rangle$~\cite{fn-large-sc}.
In all the simulations we used $N=10^5$, $\lambda=2$, 
$\kappa=10$, $k_{min}=1$ and the averages where done over $10^3$ realizations on the GC 
of networks of size $\simeq 0.6N.$ \label{Fig.2a}}
\end{center}
\end{figure}

In Fig. \ref{Fig.2a} a), we plot $\langle M_p(T,s_c)\rangle$ 
to explore the effect of $s_c$ on this average.
We can see from the plot that only for larger $s_c$ (for our simulation parameters
$\approx 200$) the curves of $P_\infty(T)$ and $\langle M_p (T,s_c)\rangle$ coincide 
for $T > T_c$ $( T_c \approx 0.34 $for$ N = 10^5)$, while for smaller $s_c$ values they do not. 
The need to use large $s_c$ to approach $P_{\infty}(T)$ had been realized previously~\cite{Newman,Eben}, but not been commented on in any detail.
We can see this behavior more clearly in Fig.\ref{Fig.2a} b), where we plot 
$P_{\infty}(T) - \langle M_p(T,s_c)\rangle $ for different values of $s_c$
and find that $P_\infty (T)$ is an upper bound of $\langle M_p(T,s_c)\rangle$,
except for very large $s_c$ (See Ref.~\cite{fn-large-sc}). From the inset of Fig.~\ref{Fig.2a} b), we can see that the difference reaches approximately $3\%$ for large values of $s_c$. 

\begin{figure}[h,t]
\begin{center}
\includegraphics[width=12cm,height=10cm,angle=0]{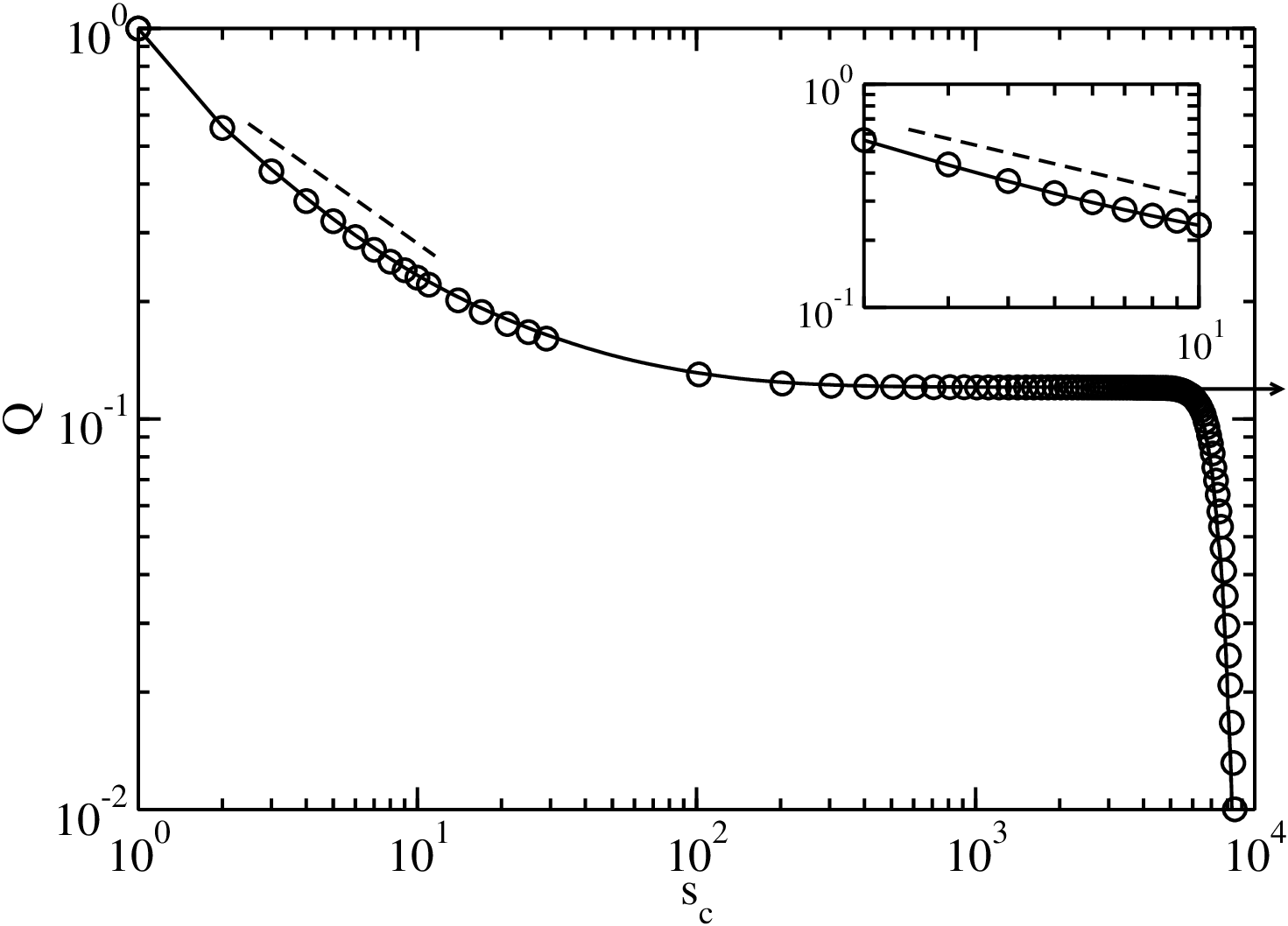}
\caption{Plot of $Q$ for: SIR as a measure of the number of times an $S_{\rm SIR}\geq s_c$ divided by the number of realizations (full line). Link percolation over all clusters as in Eq~.(\ref{Q_P}) ($\bigcirc$). We observe that both curve are in good agreement. For small $s_c$, $Q$ has a power-law decaying behavior with exponent $\tau-2=1/2$. The arrow represents approximately $S_p^\infty / \langle N_{GC} \rangle \approx 0.12$ as predicted by theoretical scaling. \label{Fig.6}}.
\end{center}
\end{figure}

The choice of $s_c$ has an extra consequence, which is to change
the likelihood that a given pathogen propagation be declared as
an epidemic. This probability is relevant from the standpoint of
readiness, because lower $s_c$ implies that it is more likely to consider almost any
disease propagation as reaching the epidemic state. 
Thus, we define $Q$ which represents the probability
that an SIR with transmissibility $T$ has size $S_{\rm SIR}\geq s_c$.
This quantity can be computed directly as the number of times 
$S_{\rm SIR}\geq s_c$ divided by
the total number of realizations (See Fig.~\ref{Fig.6}). Analytically,
$Q$ can be related to $\Phi (S_{\rm SIR})$ through
\begin{equation}
Q=\frac{\sum_{S_{\rm SIR} \geq s_c} \Phi(S_{\rm SIR})} 
{\sum_{S_{\rm SIR}\ge 1} \Phi(S_{\rm SIR})}=
\sum_{S_{\rm SIR} \geq s_c} \Phi(S_{\rm SIR}),
\label{Q_SIR}
\end{equation}
where the last equality is a consequence of normalization.
In order to calculate $Q$ from the percolation results, we
keep in mind the reweighting applied to Eq.~(\ref{Mp}). Then, $Q$ is given by
\begin{equation}
Q=\frac{\sum_{S_p \geq s_c}  S_p \; \phi(S_p) }{\sum_{S_p\ge 1} S_p \; \phi(S_p)}\;.
\label{Q_P}
\end{equation}
where $\sum_{S_p \ge 1} S_p \phi (S_p) = \langle N_{GC} \rangle$.
In Fig. \ref{Fig.6}, we plot $Q$ for SIR for $T=0.4$, ($T \gtrsim T_c $),
using direct computation and compare it with the results obtained using
Eq.~(\ref{Q_P}). We can see that the agreement is excellent.  In order to
understand the scaling behavior of $Q$, we first consider the details of
$\phi(S_p)$. From percolation theory it is known that, for $T$ close and above 
$T_c$, $\phi(S_p)\sim A S_p^{-\tau}\exp(-S_p/S_p^{\times}) +F(S_p -
S_p^{\infty})$, where $\tau$ has the mean field value $5/2$. In the last
expression, $S_p^{\times}$ is a characteristic maximum finite cluster size
which scales as $|T-T_c|^{-\sigma}$ ($\sigma=2$), 
$A$ is a measure of the relative statistical weight between the two
terms (estimated below), $F$ is a narrow function of its argument,
and $S_p^{\infty}=S_p^{\infty}(T)\equiv \langle N_{GC} \rangle P_{\infty}(T)$. 

To calculate $Q$, we use $\phi(S_p)$ and Eq.~(5), and assume the continuum limit over
$S_p$, giving
\begin{multline}
Q\sim\int_{s_c}^{\langle N_{GC} \rangle} \frac{S_p\phi(S_p)}{\langle N_{GC} \rangle} dS_p\\
\sim\int_{s_c} \frac{[A S_p^{-\tau+1}\exp(-S_p/S_p^{\times})+ S_p F(S_p-S_p^{\infty})]}{\langle N_{GC} \rangle} dS_p\\
\sim 
\left\{
\begin{array}{ll}
A\frac{s_c^{-\tau+2}-(S_p^{\times})^{-\tau+2}}{\langle N_{GC} \rangle (\tau-2)}+ 
\frac{S_p^{\infty}}{\langle N_{GC} \rangle} &\qquad[s_c\leq S_p^{\times}]\\
\frac{S_p^{\infty}}{\langle N_{GC} \rangle}&\qquad[S_p^{\times}\ll s_c\leq S_p^{\infty}]\\
0&\qquad[S_p^{\infty}<s_c],
\end{array}
\right.
\label{Q_int}
\end{multline}
where we approximated the first term of the integral by truncating the integration
at $S_p^{\times}$, and simplifying $F$ to a delta function (of integral 1, which relates to
the value of $A$). 
Several $Q$ regimes can be identified: (i) for $s_c\ll S_p^{\times}$, the contribution
of $(S_p^{\times})^{-\tau+2}$ is negligible and therefore $Q\sim s_c^{-\tau+2}$; 
(ii) for $s_c\sim S_p^{\times}$, $Q$ becomes dominated by a competition 
between the two terms of the integral and no clear scaling rules apply; 
(iii) for $S_p^{\times}\ll s_c<S_p^{\infty}$, 
$Q\sim S_p^{\infty}$, and; (iv) for $s_c>S_p^{\infty}$, $Q\rightarrow 0$.
From Fig. {\ref{Fig.6}} we can identify those four regimes. In the figure 
the arrow represents approximately  $S_p^\infty / \langle N_{GC} \rangle \approx 0.12$ 
from the simulation. The agreement between the theoretical scaling (see Eq.~({\ref{Q_int}})) 
and the simulation is excellent.

Moreover, the value of $A$ can be estimated from the fact that, for a system size
$\langle N_{GC} \rangle$, the first term of $\phi(S_p)$ accounts for the finite clusters present, and
the integral of $S_p\phi(S_p)$ must be equal to the mass of the finite clusters. Therefore
\begin{eqnarray}
[\langle N_{GC} \rangle -S_p^{\infty}(T)]\sim A\int_{1}^{\langle N_{GC} \rangle} S_p^{-\tau+1}\exp(-S_p/S_p^{\times})dS_p \nonumber \\
\Rightarrow A\sim \frac{(\tau-2)(\langle N_{GC} \rangle -S_p^{\infty}(T))}{1-(S_p^{\times})^{-\tau+2}}.
\end{eqnarray}
Since the rest of the mass of the network is contained in a single spanning cluster,
then the relative weight of the first to second term of $\phi(S_p)$
is $A:1$, justifying the choice of the integral of $F$ to be 1.
The effects shown here hold also for other networks including real networks as shown below.

One final result that can be derived from $\phi(S_p)$ is the value of $s_c$ for which 
$P_{\infty}(T)$ is a good approximation for $\langle M_p(T,s_c)\rangle$. From the
previous results, we note that there is a ``gap'' in the distribution
of sizes between $S_p^{\times}$ and $S_p^{\infty}$, which means that 
percolation generates very few clusters between these sizes. Thus, when determining
$\langle M_p(T,s_c)\rangle$, the significant statistical contributions are concentrated
in clusters smaller than $S_p^{\times}$ and then in $S_p^{\infty}$. For $s_c>S_p^{\times}$,
only the latter term contributes, driving 
$\langle M_p(T,s_c)\rangle\rightarrow P_{\infty}(T)$. It is important to recognize that
this result is independent of the system size $N_{GC}$, but not of $T$, as $S_p^{\times}$
is a function of $T$.

\section {Application to Traceroute and DIMES networks}
\label{real-nets}
The results we have presented for our model of human infectious disease propagation is
applicable to other problems in the real world. This can be well illustrated
for computer networks in which information is being broadcasted.

One of the networks that describes the functional connectivity of the Internet
is the Traceroute network, where the nodes are the routers and the
links are the connection between them that transport IP packets. The network,
as measured in Ref.~\cite{Havlin}, has $N = 222934$ nodes and $L = 279510$ links.
This network can be represented by a Scale-Free network with 
$\lambda = 2.1$~\cite{Havlin}.
In order to obtain information of the Internet connectivity, a software probe is used
called a Tracerouter tool, that sends IP packets on the Internet eliciting a
reply from the targeted host. By citing the information of the packets' path to
the various destinations, a network of router adjacencies is
build~\cite{Pastor-Satorras_book}. Here, the SIR process can be understood as a router
that has a random failure (Infected), that can produce failures on
neighbor nodes that are functional (Susceptible), and these new nodes become
infected. Thus, after some time the router is practically disconnected
from the communication network (Removed). The DIMES network ~\cite{DIMES} uses the same
algorithm of searching than the Tracerouter network, the nodes are
Autonomous Systems (AS) and the links are the connections between AS. The network
has $N = 20556$ nodes and $L = 62920$ links. The description of the SIR
process over DIMES is the same as the one explained before for the Tracerouter network.

\begin{figure}[h,t]
\begin{center}
\includegraphics[width=12cm,height=10cm,angle=0]{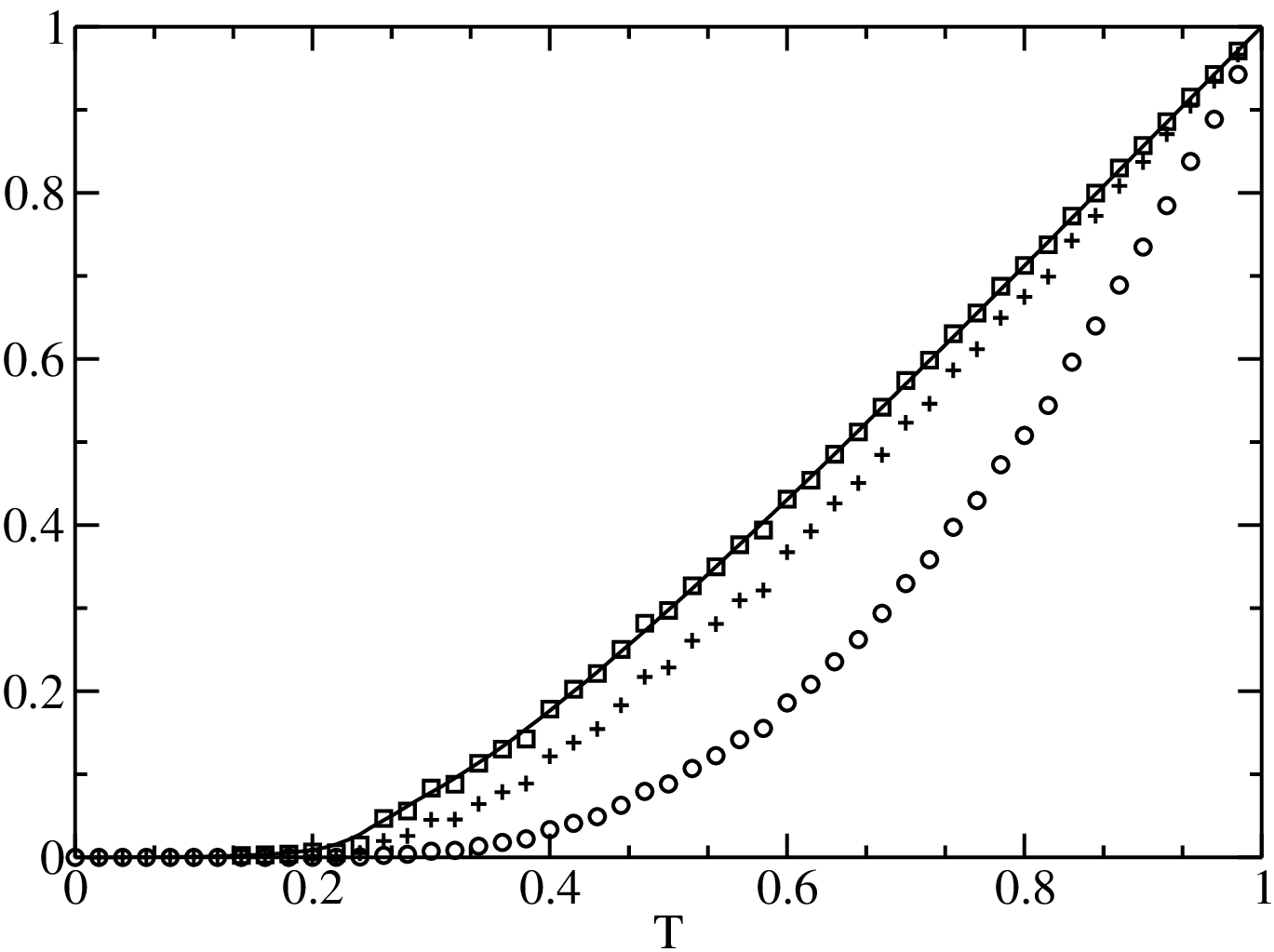}
\caption{Plot of $\langle M_{p}(T,s_c) \rangle$ as a function of $T$, for
  the Tracerouter network that has
$N=222934$, Links $=279510$, $P(k)\sim k^\lambda$ with $\lambda = 2.1$, with $s_c=0$ ($\circ$),
$s_c=2$ ($+$) and $s_c=100$ ($\Box$). The full line represents
$P_\infty(T)$.\label{Fig.4}}
\end{center}
\end{figure}

\begin{figure}[h,t]
\begin{center}
\includegraphics[width=12cm,height=10cm,angle=0]{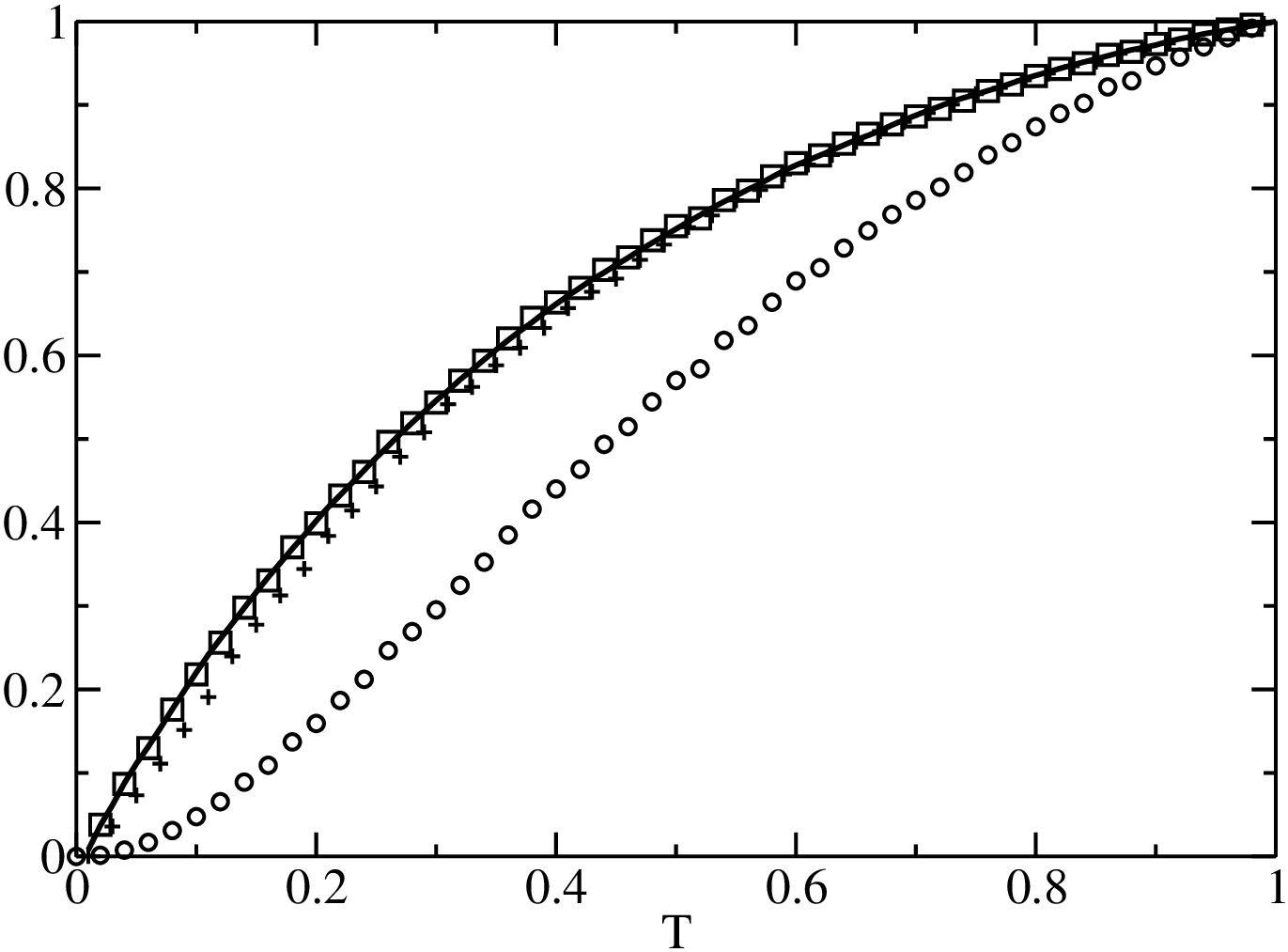}
\caption{Plot of $ \langle M_{p}(T,s_c)\rangle$ as a function of $T$,
  for the DIMES network that has Scale Free distribution with $\lambda \approx 2.15$,
  $N=20556$, Links $=62920$, for $s_c=0$ ($\circ$), $s_c=10$ ($+$) and
  $s_c=500$ ($\Box$). The full line represents $P_\infty(T)$.\label{Fig.5}}
\end{center}
\end{figure}

\begin{figure}[h,t]
\includegraphics[width=10cm,height=8cm,angle=0]{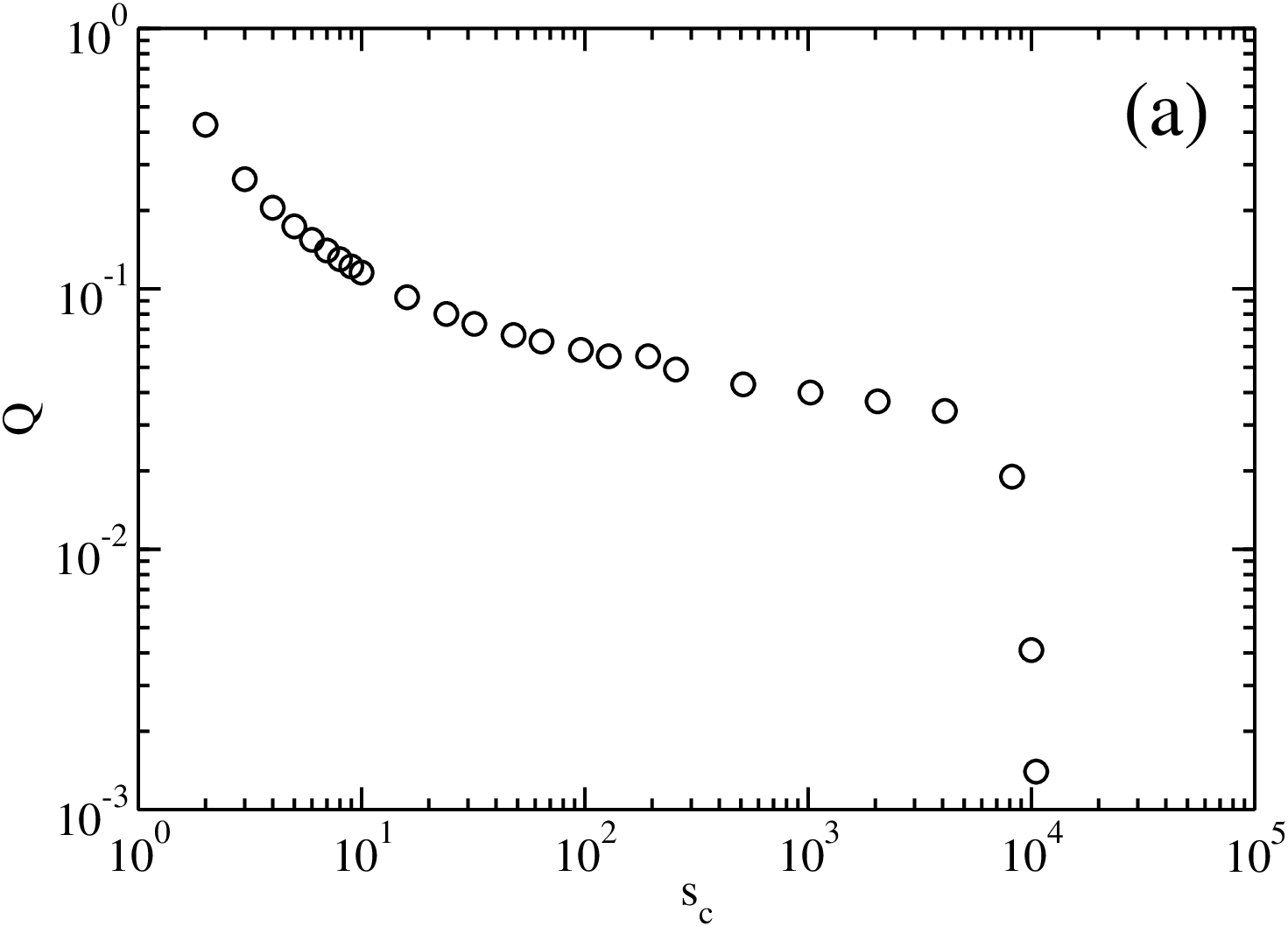}
\vspace{2cm}
\hspace{1cm}
\includegraphics[width=10cm,height=8cm,angle=0]{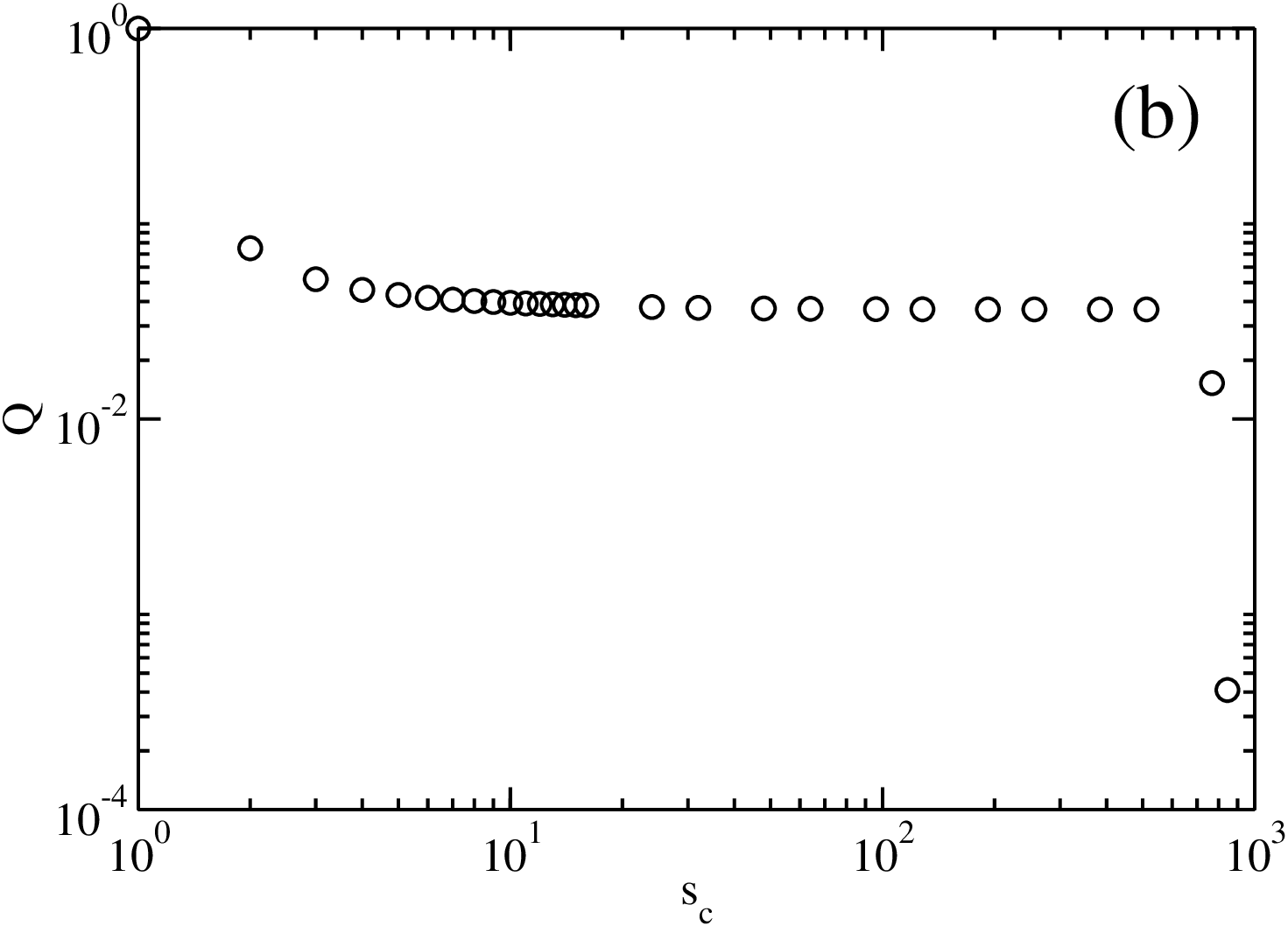}
\caption{$Q$ as a function of $s_c$ for:
\textbf{a)} Tracerouter network, with $T = 0.25$ ($\bigcirc$).
\textbf{b)} DIMES network, with $T = 0.02$ ($\bigcirc$), the exponent of the decreasing power-law is around 0.62, indicating that for this network $\tau \sim 2.62$.\label{fig.3}}
\end{figure}

In Figs.~\ref{Fig.4} and \ref{Fig.5} we plot $P_\infty(T)$ and $\langle M_p(T,s_c) \rangle$ 
for different values of $s_c$ as a function of $T$. For $s_c = 500$, for Tracerouter and
$s_c = 100$ for DIMES network we can map this problem to $P_{\infty}(T)$ of link percolation. 
We can see that the problem maps into $\langle M_p(T,s_c) \rangle$ for any size of $s_c$.
We compute $Q$ for both networks, those result are plotted in Fig.{\ref{fig.3}} a) 
and b) for Tracerouter and DIMES networks, respectively.
For DIMES, $T_c \rightarrow 0$, and thus first region cannot be seen~\cite{Stauffer}. 
On the other hand, if $T_c$ is finite as in Tracerouter, $Q$ has the four regions 
described for model networks (see Eq.~(\ref{Q_int})).

\section {Summary}
\label{conclusions}
We have shown that the choice of $s_c$, the minimum SIR propagation size
necessary to declare an epidemic, has important consequences on 
epidemiological predictions.
Using percolation theory to calculate the average fractional size 
$\langle M_{\rm SIR}(T,s_c)\rangle=\langle M_p(T,s_c)\rangle$ of an epidemic, 
we find that the strength of the spanning link percolation cluster $P_{\infty}(T)$ 
is an upper bound to $\langle M_{\rm SIR}(T,s_c)\rangle$, provided $s_c$ does not
exceed $S_p^{\infty}(T)$, the typical size of finite clusters of link percolation,
where pathological results can appear. 
When $s_c$ is between $S_p^{\times}(T)$ and $S_p^{\infty}(T)$, $P_{\infty}(T)$
is a good approximation to $\langle M_{\rm SIR}(T,s_c)\rangle$.
For small values of $s_c$, $P_{\infty}$ is no longer 
a good approximation, and the average fractional size has to be computed directly. 
We also study $Q$, the probability that an SIR propagation reaches
the epidemic mass $s_c$, which has several interesting regimes including one that
scales as $s_c^{-\tau+2}$.
We apply our results to real networks (DIMES and Tracerouter) to measure the
consequences of the choice $s_c$ on predictions of average outcome sizes of computer
failure epidemics.

\subsubsection*{Acknowledgments}

E.L. acknowledges financial support from DOE (US) and EPSRC (UK). 
C.L., M.V.M., P.A.M. abd L.A.B acknowledge financial support from 
PICTO-3370 (ANPCyT) and U.N.M.d.P. We also acknowledge M. E. J. Newman 
and  E. Kenah for fruitful discussions.

\vspace{1.5 cm}

\vspace{1cm}

\vspace{1 cm}

\newpage

\begin{thebibliography}{99}
\bibitem{fn-polylogarithm} The function $\mbox{Li}_{\lambda}$ is the polylogarithm function
of argument $\lambda$, which emerges in this context as a consequence of the normalization
condition of the probability distribution $P(k)$.
\bibitem{fn-uncorrelated-networks} This approach is valid when, statistically,
the number of links of a given node has no correlations to the number of links
of its neighboring nodes.
\bibitem{algorithm} In this algorithm~\cite{Molloy}, each node is first assigned a 
random number
of ''stubs'' taken from $P(k)$. Next, we connect two unused stubs from two randomly
selected nodes. The only condition that we impose is that there cannot be multiple
edges between two nodes.
\bibitem{fn-time} In this article, since overall SIR dynamics
are not considered, $t$ refers to the ``local'' time, i.e., the time of
every individual node. Thus, $t=0$ refers to the time of infection of a given node.
More formally, one could refer to $t_i$ for the time running for node $i$, but by context
it is clear that this is not necessary.
\bibitem{fn-allclusters} Note that Eqs.~(\ref{MSIR}) and (\ref{Mp}) involve
also the largest cluster. For model networks of finite size (see Eq.~({\ref{PdK}})) 
and for real networks, the size of the largest cluster does not diverge.
\bibitem{fn-large-sc} When $s_c$, and given that $\langle M_p(T,s_c)\rangle$ only
counts $S_p\geq s_c$, it is possible to have $P_{\infty}<\langle M_p(T,s_c)\rangle$,
because averaging of the mass is only taking place in those rare realizations
when the mass condition is satisfied. In this case, one can see a violation
of $P_{\infty}(T)$ being a bound for $\langle M_p(T,s_c)\rangle$, but this
behavior is pathological.
\end{thebibliography}
\end{document}